\documentclass[aps,prab,reprint,superscriptaddress,showpacs,showkeys,letter]{revtex4-1}
\usepackage{graphicx}

\usepackage{dcolumn}
\usepackage{bm}
\usepackage{amsmath}
\usepackage[caption=false]{subfig}
\usepackage[none]{hyphenat}
\usepackage{color}
\usepackage{mathrsfs}
\usepackage{epstopdf}
\usepackage{amssymb}
\usepackage{mathrsfs}

\begin{document}


\title{Interaction between high intensity beam and structure resonances}

\author{Chao Li}
\email{lichao@ihep.ac.cn}
\affiliation{%
Key Laboratory of Particle Acceleration Physics and Technology, Institute of High Energy Physics, Chinese Academy of Sciences, 19(B) Yuquan Road, Beijing 100049, China.}

\author{R. A. Jameson}
\affiliation{%
Institut f\"{u}r Angewandte Physik, Goethe Uni Frankfurt, Frankfurt-am-Main, Germany.}


\begin{abstract}
With the help of the linearized perturbation theory, the collective beam instability due to the space charge, modelled by the Vlasov-Poisson equation, is well studied in the former research theoretically in a general sense  [Chao Li and R. A. Jameson PRAB, 21, 024204 (2018)], where the related instability is depicted by the ``structure resonance". As an extension of the theoretical study,  this paper aims to verify the validity of the structure resonance with numerical simulations and  to show the basic principles which interpret the interaction between structure resonance and beam in a general sense. It is found that  the  reaction of beam to the structure resonance ($n\sigma\approx 180^\circ$) differs for the cases with $n= 0$  and $n\ne 0$ stop bands. The non-resonant ``plasma oscillation" is also revisited; the ``phase space mixing", lying in the transient plasma oscillation, is the mechanism that ties  the  resonant effects, non-resonant effects and macroscopic beam phenomena together. 

\end{abstract}


\maketitle

\clearpage

\section{Introduction}

A fundamental understanding of the equilibrium and the stability properties of high intensity beam in accelerators is crucial for the development of advanced particle accelerator applications. From the theoretical point of view, the stability study of an high intensity beam requires a self-consistent treatment of the coupled Vlasov-Poisson equations.  In the former study~\cite{1}, with an unsymmetric 4D Kapchinskij-Vladimirskij (KV) initial beam assumption \cite{5} in Alternating Gradient (AG) periodic focusing channels,  a group of equations  modelling the beam collective motions ($I_{j;k,l}(s)$ and its eigenphase $\Phi_{j;k,l}$ in one focusing period) has been successfully established with the help of the linearzied perturbation theory. The  $structure$ $resonance$ ($n\sigma\approx 180^\circ$) is used to depict the resonance condition leading to a collective beam instability. If the focusing channel is smoothed by the constant focusing assumption, the structure resonance is degenerated to the resonances studied by Hofmann in Ref.~\cite{2}.

Care has to be taken in extending the  prediction from structure resonance based on the 4D KV beam to other beam distributions. The key point is that the density discrepancy between the ideal KV beam and the real beam distribution plays both the roles of structure resonance driving forces and resonance de-tuning (or resonance damping) force. Thus, the time scales to clearly show evidences of the structure resonance for different distribution types are different in numerical simulations. In this paper, guided by the prediction shown in Ref.~\cite{1}, we aim to give a  physical picture of the interaction between the structure resonance and high intensity beam through numerical simulations. Rms equivalent initial beam profiles, such as WaterBag (WB) and KV,  are used to consolidate the principles summarized from the  numerical observations. With various lattice conditions, the beam will be initially put into a certain structure resonance stop band to see how it will evolve in the following AG focusing periods. Using the principle of rms equivalent input distributions \cite{21}, results from different initial distributions do not show qualitative discrepancies, except in the time scales when resonance effects are observable in phase space. It is found that in the $n\ne 0$ structure resonance stop band, the regions of rms beam emittance growth, beam halo and beam loss taking place obtained from Particle-In-Cell (PIC) simulations or from experiment  in Paul Trap~\cite{3} agree reasonably well with the theoretical prediction. If a beam starts in the structure resonance stop band, in general, it will gradually  move out of the structure resonance  stop band spontaneously and get to a transient equilibrium state.  Whereas in the $n=0$ structure resonance stop band  near the space charge limit where the depressed phase advance is close to $0^{\circ}$, only the resonances $\Phi_{j;k,l}=0^{\circ}$ exist \footnote{These structure resonances actually are exactly those kept with smoothing approximation in Ref.~\cite{2}}, the beam does not evolve out of, but stays in the structure resonance stop band  for a long time  with a tiny  linear emittance growth tendency.

For a profound understanding accordingly, it is the underlying non-resonant  phase space mixing due to the space charge variation according to the varying transport  channel conditions  that ties all the beam dynamic phenomena together. We remark that this ``phase space mixing due to space charge" is the very basic mechanism for energy transfer, changing the particle distribution, and can be transiently described (as function of time or position) by the transient plasma period $\omega_p$ -- a transient description \cite{4}.

It is noteworthy that the solvable linearized Vlasov-Poisson system is limited to the 4D case.  Actually, although only the coasting beam in rings or the beam in Paul Trap with a small longitudinal oscillation frequency can approach this 4D condition; to analyse the bunched beam, the  described mechanism of instability also closely describe the longitudinal-transverse space charge coupling cases in ion accelerators. In section II, we will briefly introduce the basic numerical approaches and physical terminologies used in  this manuscript. In addition, two examples, comparison between the results from Paul Trap experiments and the structure resonance prediction, and comparison between the PIC simulation and structure resonance prediction,  will be given to demonstrate the validity of the structure resonance mechanism. In section III, firstly, a basic set of principles for the interaction between beam and  structure resonance ($n\ne0$) is summarized; then one typical example is given to show how these principles are used as interpretations in simulation observations.  Most important, we emphasize  the extended view  of the space charge beam physics, showing the complex time dependent process and phase space transport in  a transient sense, beyond the usual steady state presentation. In section IV, we extend the beam evolution discussion  to the space charge limit $\sigma \approx 0^{\circ}$. The resonant structure resonance scattering, leading to linear emittance growth tendency, and non-resonant beam oscillation within plasma period $\omega_p$, leading to the ``particle redistribution", are explained respectively.  We state that the ``phase phase mixing",  lying in this transient plasma oscillation, is the mechanism that  ties the  resonant effects, non-resonant effects and macroscopic beam phenomena together. The summary and discussions are given in section V. 

\section{Basic physical model, numerical approaches and  validity of the structure resonance}
We briefly introduce the basic physical model of structure resonance. The study of the collective beam instability has to ensure the self-consistence, which requires to solve the coupled Vlasov-Poisson equations below,  
\begin{eqnarray}\label{eq2.3}
  &&\frac{\partial f_1}{\partial s} + [f_1,H_0] +  [f_0,H_1]=0,   \nonumber \\
  &&\Delta H_1(x,y) = \frac{1}{\epsilon_0}  \int \int f_1 dxdy.
\end{eqnarray}
where ($H_0$, $f_0$) and ($H_1$, $f_1$) are the ideal and space charge perturbed Hamiltonian and distribution function respectively. In general, $H_1$ can be expressed as a polynomial inside near the beam boundary. As to different orders of the polynomial expressed, various  beam collective modes can be established. Inheriting the terminologies defined in Ref.~\cite{1,16},  the collective modes, $I_{j;k,l}(s)$, represent integrals of the surface electric field discontinuity from period to period; the phase advance of $I_{j;k,l} (s)$ is noted as $\Phi_{j;k,l}$;  $\sigma_s$ is used to describe the single particle phase advance. The nonlinear effects from external elements and internal space charge cause different particles to have different particle phase advance $\sigma_s$, leading to a beam phase advance spread; $\sigma_0$ and $\sigma$ are standard notation used to evaluate the average focusing strength in one focusing period without and with space charge.

For clarity, we limit the study to the periodic AG channel where only the transverse beam motions are taken into account. The matched beam sizes in a FD focusing channel meet the conditions $a(s)=a(s+S)$ and $b(s)=b(s+S)$, where $a(s)$ and $b(s)$ are the beam size in transverse x and y direction, $S$ is the period length.  The notation $\Phi_e$ is used to depict the phase advance of the rms matched envelope oscillation characteristics in one period in this paper, which is always $360^{\circ}$.  The $structure$ $resonance$, used to depict the resonance condition leading to a collective beam instability, is  composed of $parametric$ $resonance$ with the  condition $\Phi_{j;k,l}/\Phi_e=m/2$ and $confluent$ $resonance$ with the condition $\Phi_{j;k,l}^{(1)}+\Phi_{j;k,l}^{(2)}=m \times 360^{\circ}$.  Due to the basic FD lattice  period used for study,  in the following discussion, the index $m=1$ is adopted. Thus, the notation $ n\sigma\approx \Phi_{j;k,l}\approx 180^\circ$ will be  used to roughly depict the $nth$ order of structure resonance in the AG focusing channels for simplicity.



In the following, the PIC code PTOPO \cite{8,9} is used for numerical simulations. Initially well rms matched 150 KeV proton beam, composed of 20 K macro-particles with a normalized emittance $\epsilon_x=\epsilon_y=0.25$ $cm$ $mrad$, is used as the initial condition. The Dirichlet boundary condition  is adopted for the space charge potential calculation on the surface of a rectangular pipe. 
Besides the rms quantities normally used, the depressed phase advances $\sigma$  in each period are also obtained with the help of the definition $\sigma_{x(y)}=  \int^S_0 \epsilon_{x(y)}/R(s)^2ds$, where $R(s)=a(s)$ or  $R(s)=b(s)$ is the beam rms size in x and y direction respectively. Below, we give two examples to demonstrate the validity of the structure resonance.

\subsection{Structure resonance prediction and experiment on beam losses study in Paul Trap}\label{sec2.1}

\begin{figure}[t]
  \centering
  \includegraphics[width=0.8\linewidth]{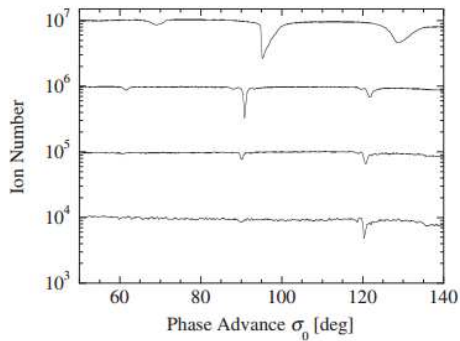}\\
  \caption{\label{fig:fig1} Ion-loss distributions measured in SPOD with the symmetric FD potential. Each of the four curves consists of 237 data points obtained from 237 independent measurements over the range $50^{\circ} <\sigma_0 < 140^{\circ}$ at the same initial ion number ($N_{ion} \approx 10^4, 10^5, 10^6$, or $10^7$). Figure is provided by the courtesy of H. Okamoto, more details can be found in Ref.~\cite{3}. }
\end{figure}

As an application of the Paul Trap equipment, Simulator of Particle Orbit Dynamics (SPOD) \cite{3} provides a non-neutral plasma which can be   physically scaled  to a charged particle beam in a AG focusing channel. In contrast with conventional experimental approaches relying on large-scale machines, it is straight forward in SPOD to control the doublet geometry characterized by the quadrupole filling factor and drift-space ratio. Fig.~\ref{fig:fig1} shows the beam loss experiment in SPOD launched in  H. Okamoto's group in Hiroshima University \cite{3}. The 4 curves represent different beam intensity settings. For each curve, the beam intensity is fixed and the external focusing (the zero beam current phase advance $\sigma_0$)  is scanned from $50^{\circ}$ to $140^{\circ}$. Thus, the tune depression is not fixed during the each scanning. The depths of ``dips" in the curves found in experiment represent the relative beam loss due to  resonances. In the flat region, there is no beam loss taking place. More recent results and experimental details can be found in Ref.~\cite{3}.  

\begin{figure}[t]
  \centering
  \includegraphics[width=0.8\linewidth]{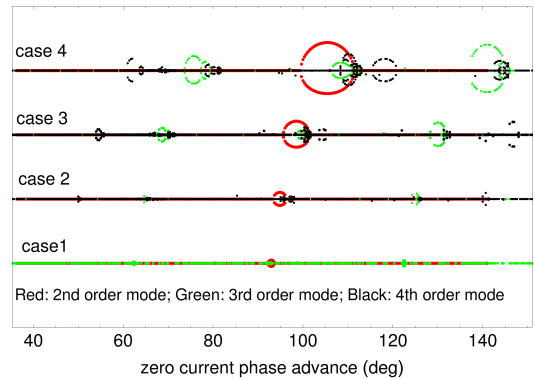}\\
  \caption{\label{fig:fig2} The structure resonance stop bands of 4 separated cases; the initial rms beam emittances are fixed with the same value and the beam currents are scanned as  $I_0$, $2I_0$, $4I_0$ and $8I_0$ from Case 1 to Case 4.}
\end{figure}

As a comparison, Fig.~\ref{fig:fig2} shows the width and the strength of the structure resonances given by the analytical model for 4 cases, with different beam currents respectively, as a function of zero beam current phase advance $\sigma_0$.  In each case,  these circles represent beam collective instability growth rate due to structure resonance and different colours represent  different orders of structure resonances, when the red, green and black show the 2nd, 3rd and 4th order structure resonance respectively.  The distance from the circle to the flat line represents the growth rate of a specific mode. The largest growth rate value in Case 4 belongs to the 2nd order structure resonance with a value around 0.22. The growth rate of other structure resonances can be  linearly scaled.  With beam current increasing (below to above), the growth rate for each order of structure resonance is getting larger, the width of the stop band is getting wider, the centres of the stop bands increase.   

The initial beam rms emittance changes  depending on the number of ions stored in the SPOD \cite{3}, thus the analytical structure resonance study here does not exactly correspond to the experiment condition. However,  compared with the ``dip" positions  in Fig.~\ref{fig:fig1}, the movement tendency of the width and the centre position as function of  beam intensity is consistent with structure resonance prediction. Roughly, the tendency of the loss as function of the beam current is also comparable with the growth rate predicted from structure resonance.   

\subsection{Emittance growth in structure resonance stop bands with simulations}\label{sec2.2}

\begin{figure}
    \centering
    \subfloat[\label{fig12:subfig:a}]{
        \includegraphics[width=.8\linewidth]{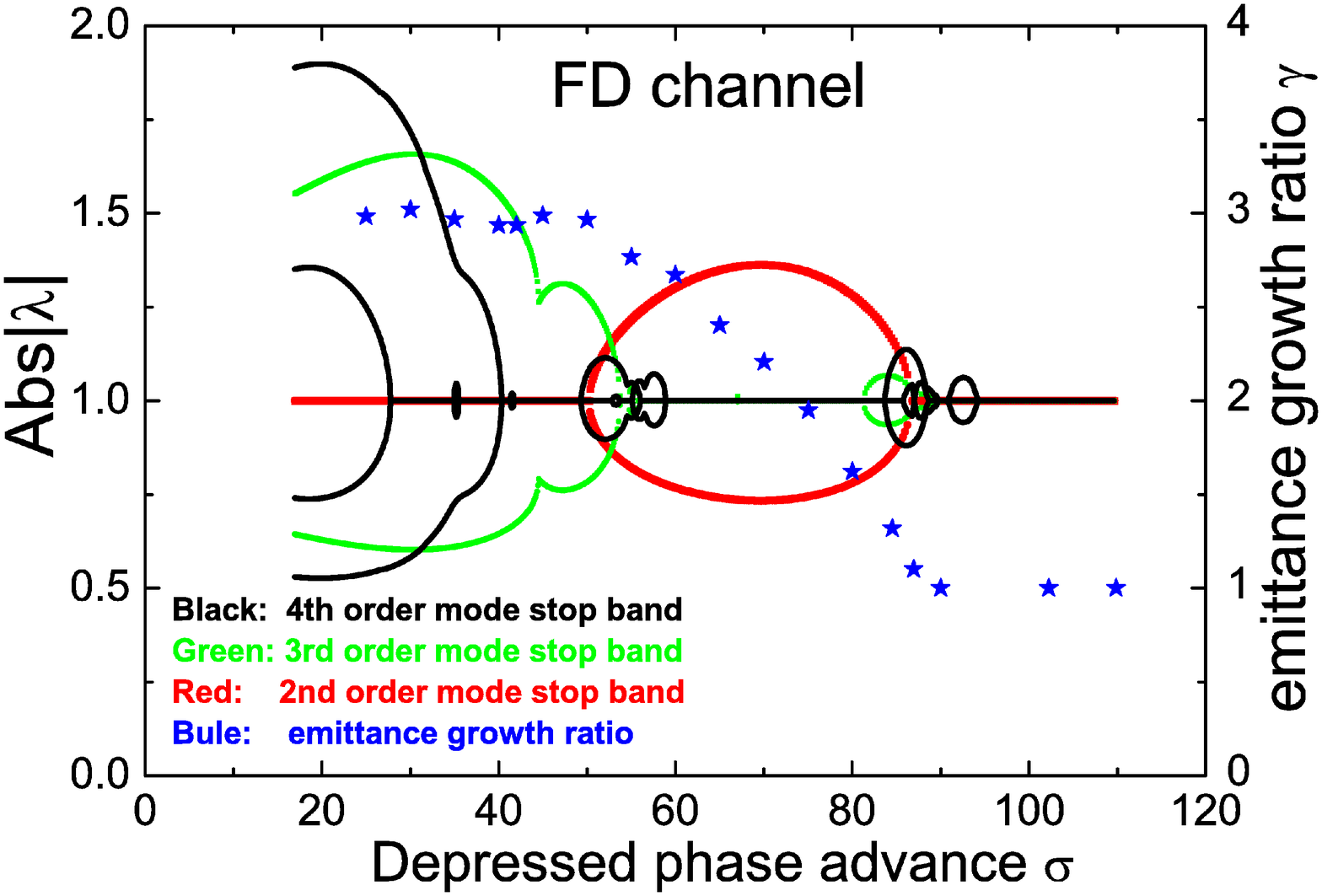}
    }
    
    \subfloat[\label{fig12:subfig:b}]{
        \includegraphics[width=.8\linewidth]{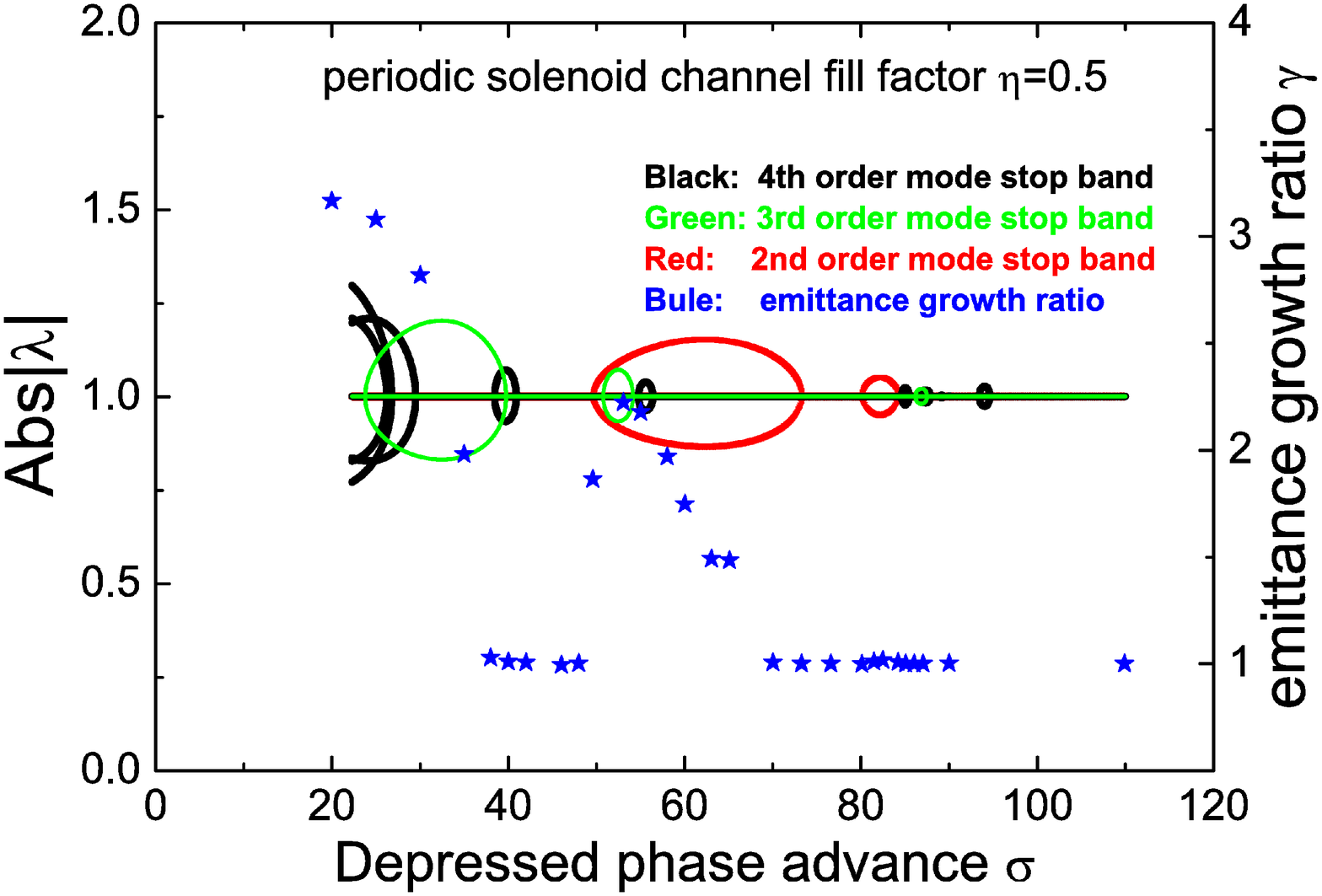}
    }
    \caption{\label{fig:fig12}Eigenvalues  of the 2nd order even (red), 3rd order (green) and the 4th order (black) structure resonance;  emittance growth ratio (blue) $\gamma$ as function of depressed phase advance $\sigma$ for a FD channel (above) and a interrupted SOLENOID (below) channel with the zero beam current phase advance $\sigma_0=110^{\circ}$. The beam is initially rms matched. }
\end{figure}

Figure~{\ref{fig:fig12}} shows the different orders of structure resonance stop bands with the condition that the zero beam current phase advance $\sigma_0=110^{\circ}$. Each of the blue stars is the emittance growth resulting from a PTOPO simulation in which the beam injected with a current giving the indicated $\sigma$.  To avoid a artificial rms emittance growth in the projected beam profile, especially in the solenoid which linearly couples the particle motions in $x$ and $y$ together, the growth ratio of the total 4D beam emittance  in the phase space $(x,p_x,y,p_y)$ is used as an evaluation for structure resonance. The total number of focusing periods is chosen large enough to ensure the emittance growth is saturated. In an anti-symmetric FD focusing channel, Fig.~\ref{fig12:subfig:a}, the emittance growth ratio $\gamma\ne1$ only when the beam is initially injected into the structure resonance stop bands. There is no stable area when the depressed phase advance $\sigma$ is less than around $90^{\circ}$ because the 2nd, 3rd and 4th orders of structure resonance stop bands are overlapped. In the interrupted solenoid channel, Fig.~\ref{fig12:subfig:b},  there is one stable gap between  $40^{\circ}< \sigma<50^{\circ}$  in which the beam emittance almost does not increase at all, since the higher order structure resonances are clearly separated.

\section{Beam interaction with the structure resonance $ n\sigma\approx \Phi_{j;k,l}\approx 180^\circ$, $n\ne 0$ }

The interaction between structure resonance ($n\sigma\approx \Phi_{j;k,l}\approx 180^\circ$, $n\ne 0$) and  beam has been studied for quite a while by different researching groups \cite{3, 10,11,17,18, 4}. As one  example, the 2nd order  structure resonance, which describes the same physics as the well-known envelope instability, occurring around $\sigma\approx 90^{\circ}$ phase advance has been well studied both theoretically and experimentally \cite{12,13,14,15}. Summarizing recent researches on this subject, we conclude some basic principles to understand the interaction between beam and structure resonance: 
\begin{enumerate}
\item The interaction between beam and structure resonance has to be studied in a transient sense, representing by transient phase advance -- transient description. 
\item Low order structure resonance stop band is a component of the higher order stop band \cite{8}.  Excluding the lower order one, pure higher order structure resonances are usually weak and can be easily washed out by the density nonlinearity. 
\item The above threshold of the structure resonance stop band has an ``attracting" effect, whereas the below  threshold has a ``repulsive" effect. 
\item The n-fold phase space structure is the direct evidence for the nth order structure resonance. 
\item The nonlinarity from density mismatch   plays  both the roles of structure resonance driving source and resonance detuning force.
\item Under the influence of the structure resonance, the beam will spontaneously evolve to the resonance free region, where the space charge takes a weaker importance with (example dependent) emittance growth and halo generation. 
\item  In this example with a long, repeating focusing channel, the beam will finally reach to a ``quasi-equilibrium" state with emittance growth saturated, which is a comprise of structure resonance and the structure resonance de-tuning.
\item  Without lattice imperfection and rms beam mismatch, the incoherent particle-core resonance is not the dominant mechanism to beam halo \cite{8}. 
\item The interaction between particle $\sigma_s$ and collective modes $\Phi_{j;k,l}$ extends the halo description to a 4D action-angle frame in a general sense. 
\end{enumerate}   

\begin{figure}[tbp]
  \centering
  \includegraphics[width=1\linewidth]{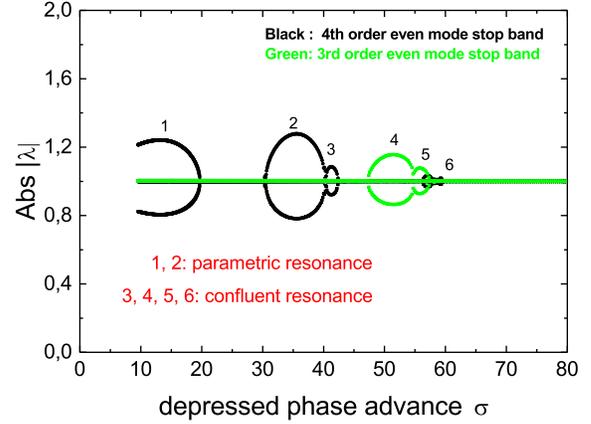}\\
  \caption{Structure resonance stop band in FD focusing channel when $\sigma_0=80^{\circ}$. }\label{fig:fig17}
\end{figure}

\begin{figure*}[htbp]
    \centering
    \subfloat[\label{fig25:subfig:a}]{
        \includegraphics[width=.45\linewidth]{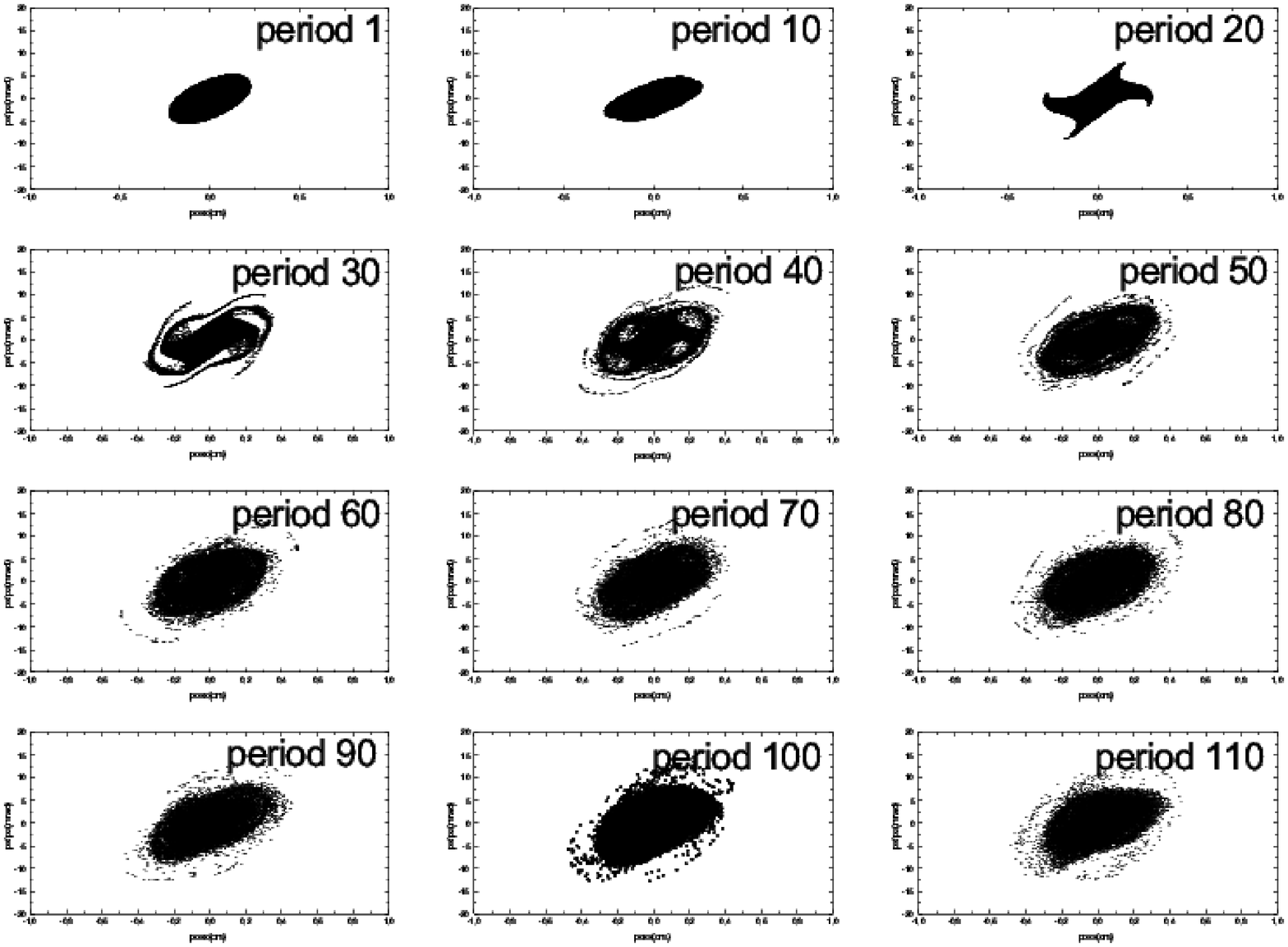}
        \includegraphics[width=.45\linewidth]{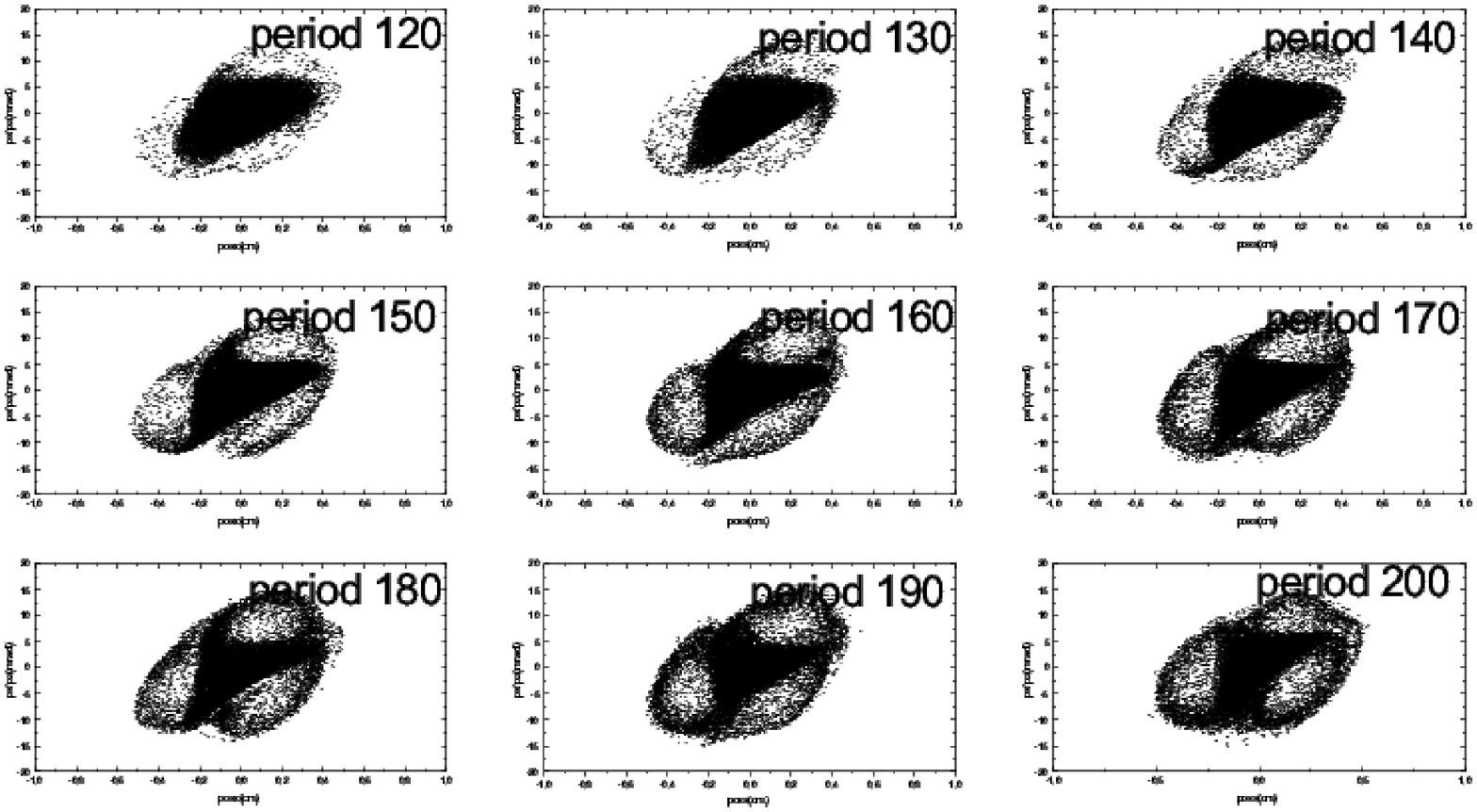}
    }
    
    \subfloat[\label{fig25:subfig:b}]{
        \includegraphics[width=.45\linewidth]{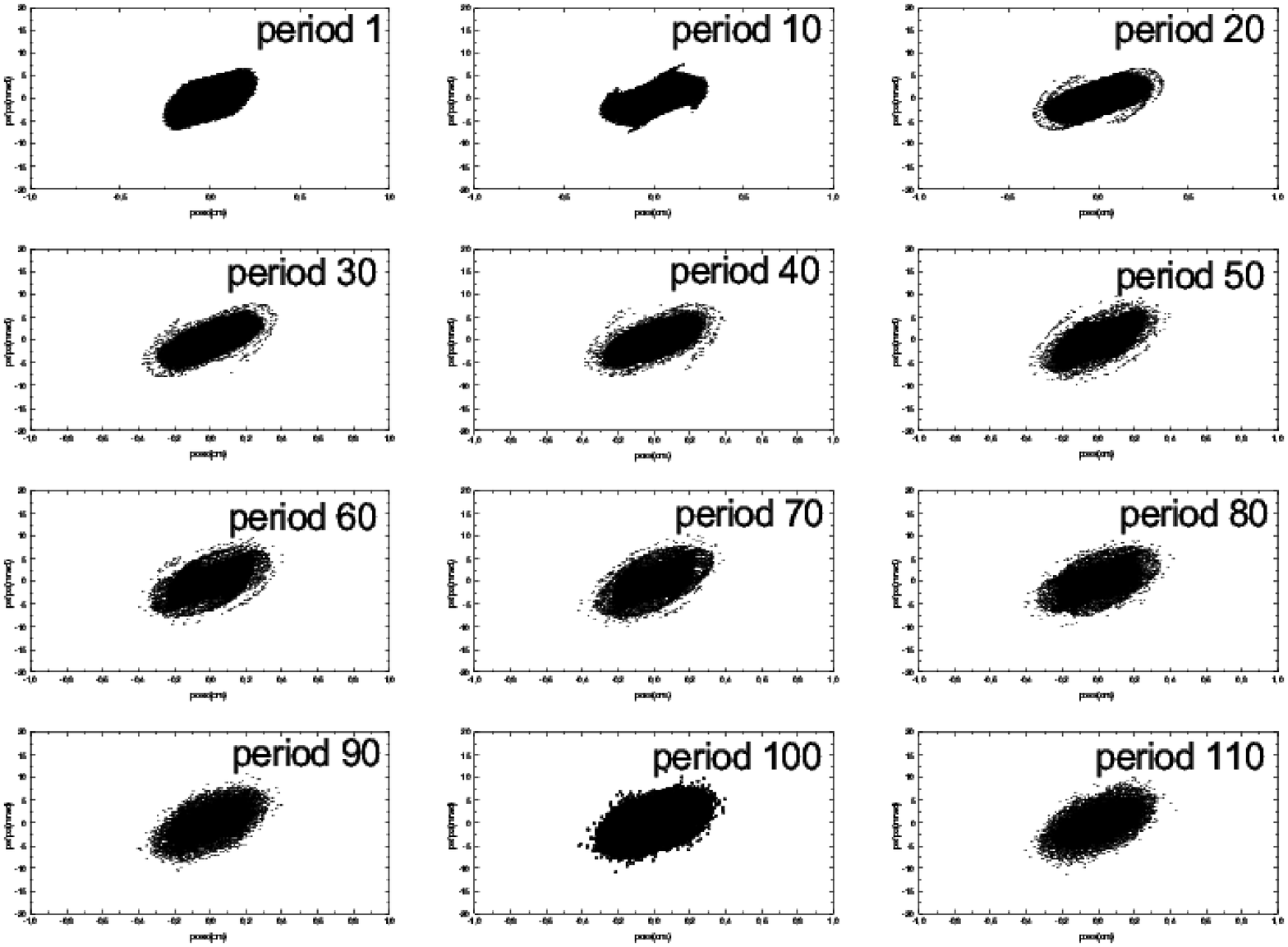}
        \includegraphics[width=.42\linewidth]{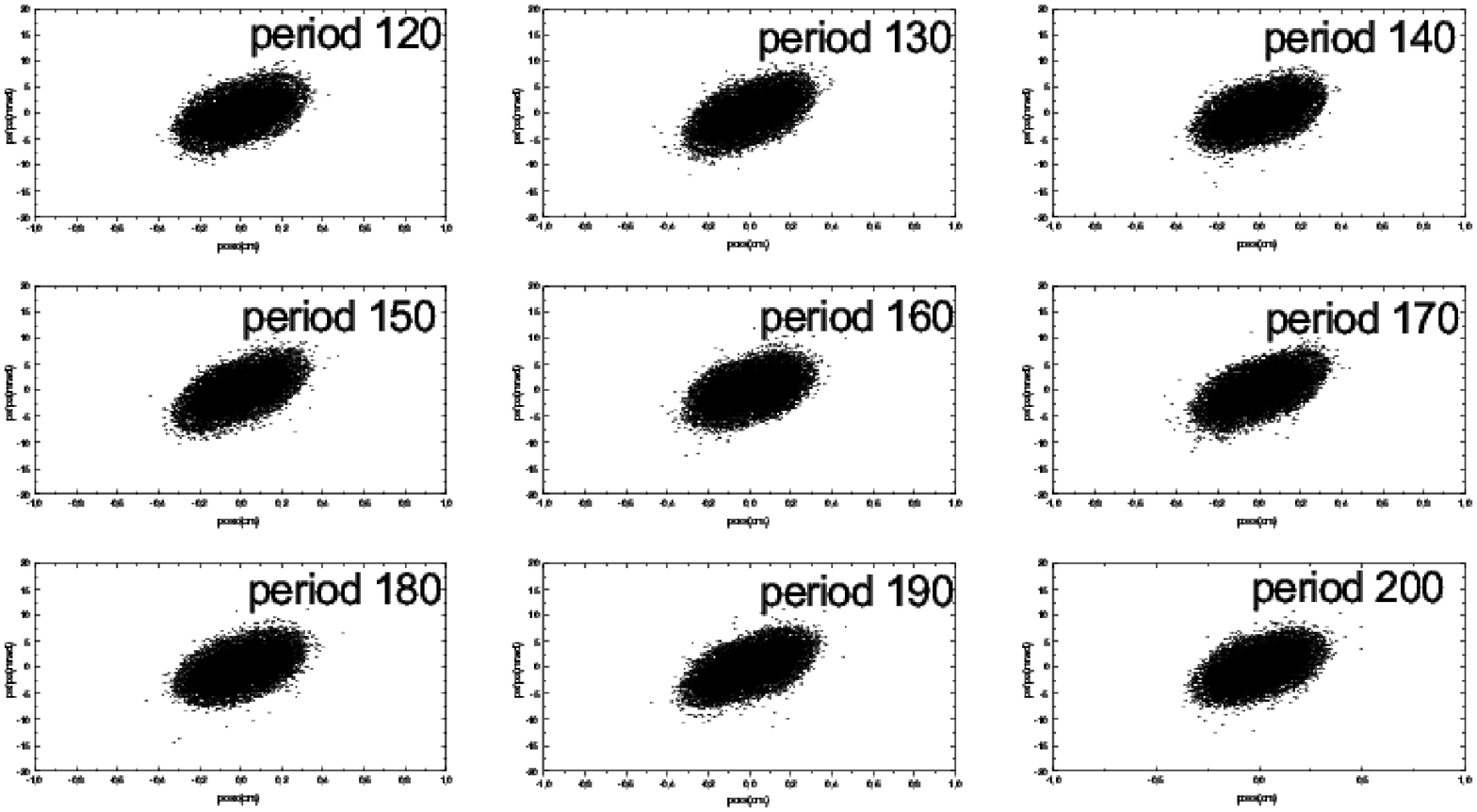}
    }
    \caption{\label{fig:fig25} Beam phase space evolution with a KV (a) and WB (b) initial beam in FD channel when $\sigma_0=80^{\circ}$, initial depressed phase advance $\sigma=35^{\circ}$. }
\end{figure*}

\begin{figure*}[t]
    \centering
    \subfloat[\label{fig23:subfig:a}]{
        \includegraphics[width=.4\linewidth]{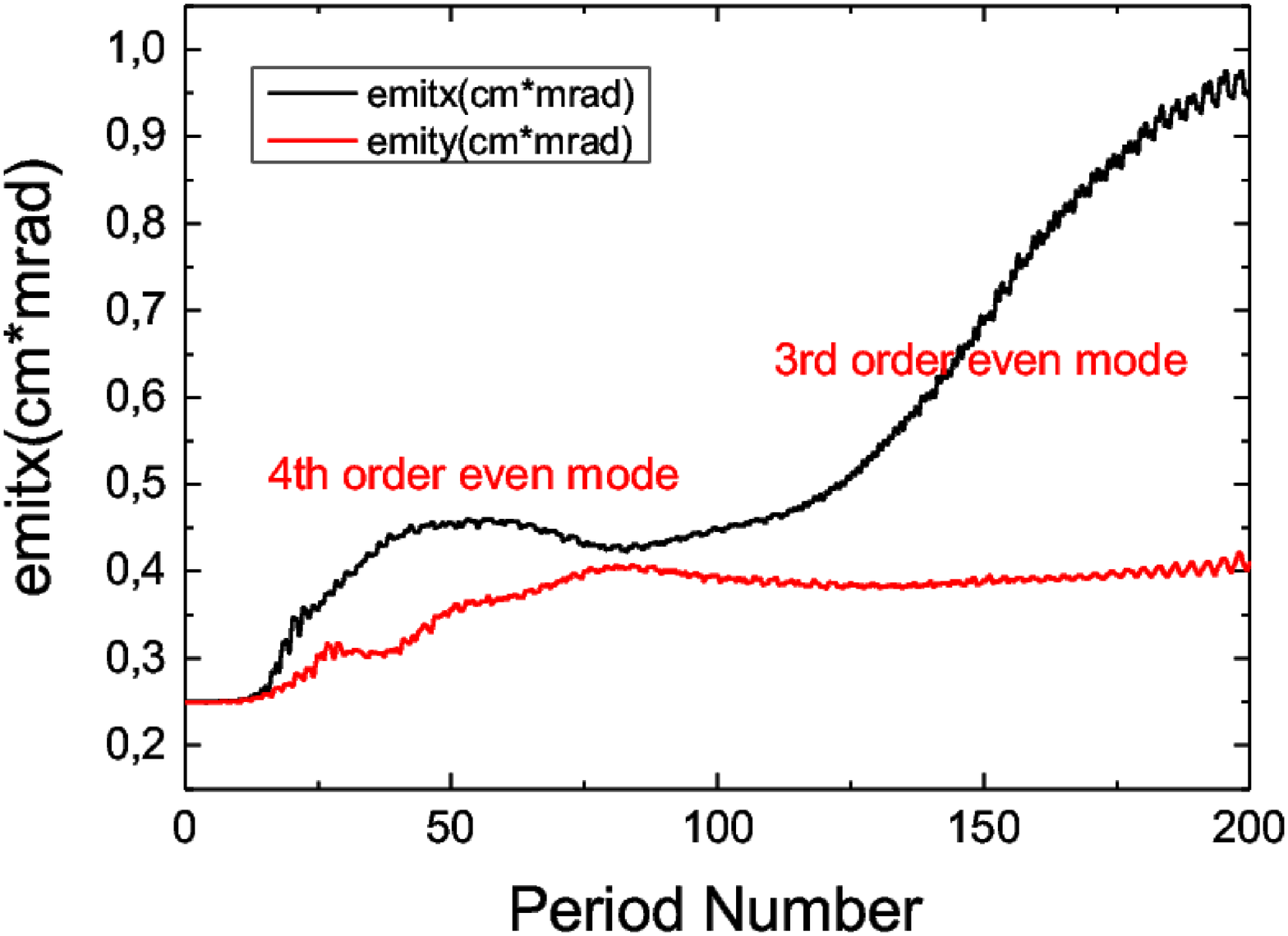}
    }
    \subfloat[\label{fig23:subfig:b}]{
        \includegraphics[width=.4\linewidth]{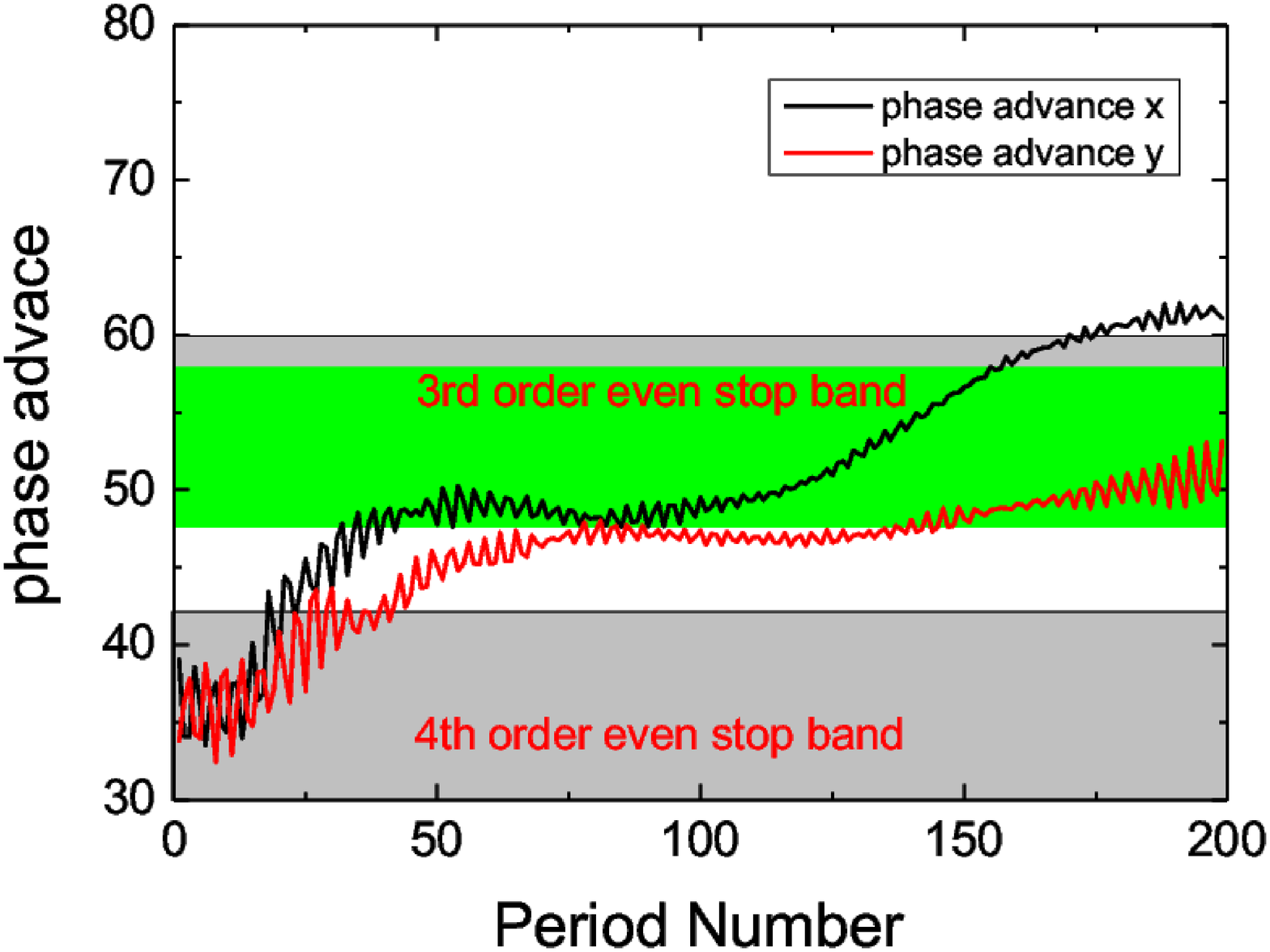}
    }
    
    \subfloat[\label{fig23:subfig:c}]{
        \includegraphics[width=.4\linewidth]{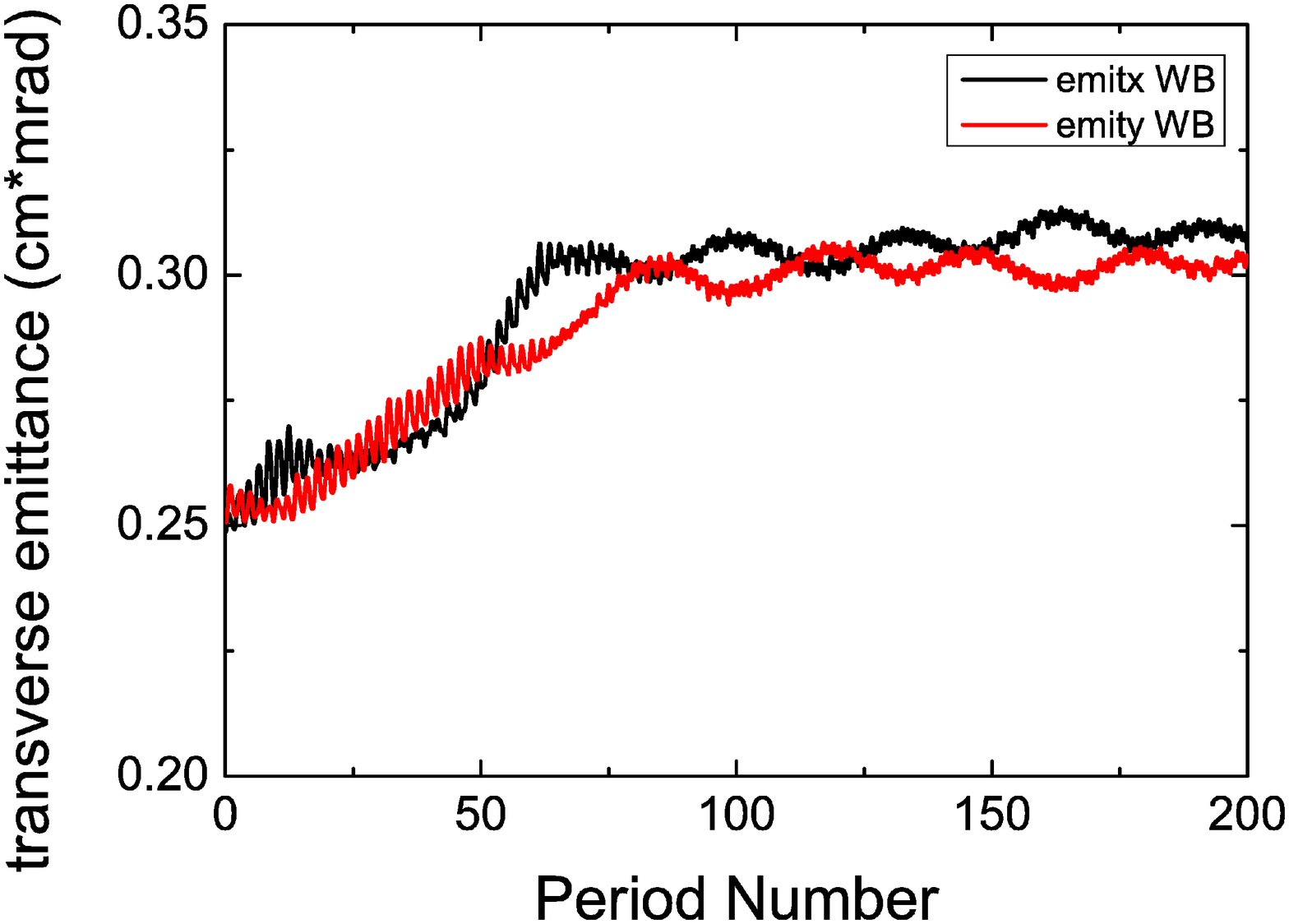}
    }
        \subfloat[\label{fig23:subfig:d}]{
         \includegraphics[width=.4\linewidth]{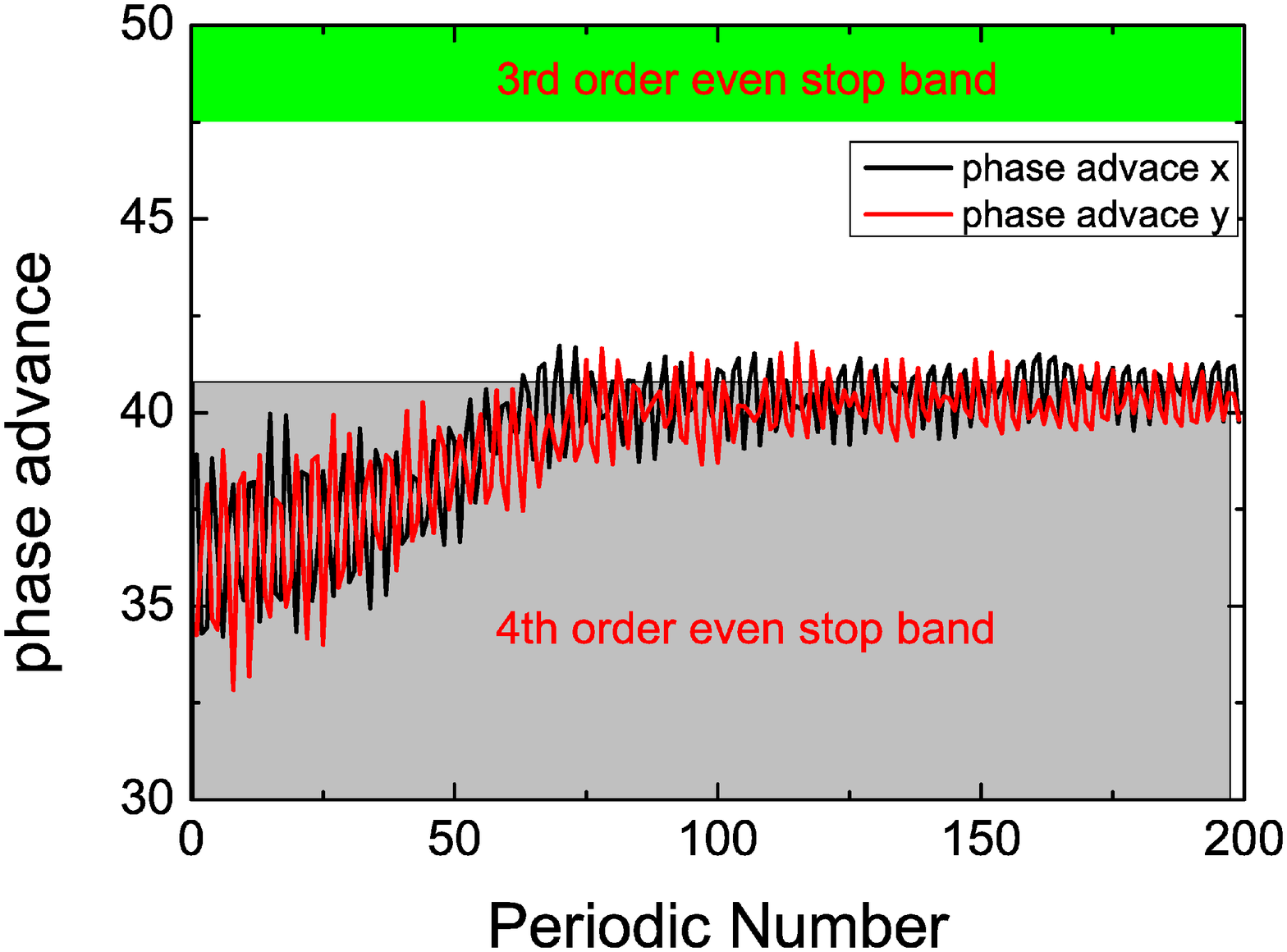}
    }
    \caption{\label{fig:fig23} The rms beam  emittance (left)  and phase advance (right) evolution in FD with a  KV (above) and WB (below) initial beam when phase advance $\sigma_0=80^{\circ}$, initial depressed phase advance $\sigma=35^{\circ}$. }
\end{figure*}

Fig.~\ref{fig:fig17} obtained from the Vlasov-Poisson model gives the 3rd and 4th order structure resonance stop band when the zero beam current phase advance $\sigma_0=80^{\circ}$. In the following, we will chose one typical case to show the influence of the  high order structure resonance (limited in the case $n\ne 0$) and to demonstrate how these  basic principles discussed above can be applied to interpret the simulation results. More simulation studies for various cases can be found in the Ref.~\cite{4}.    With the initial depressed phase advance $\sigma=35^{\circ}$ in the 4th order structure resonance stop band, Fig.~\ref{fig:fig25}  shows the phase space evolution along the channel with initial KV and WB beam distributions. Fig.~\ref{fig:fig23} depicts the  evolution of the beam emittance and the transient phase advances. 

For the initial KV beam, the perturbation  comes from noise and takes several periods to develop. The first jump of emittance growth before period 80 lies in the 4th order structure resonance. The 4-fold  beam phase space structure  at period 20 clearly shows the influence of the 4th order structure resonance. Because there is not enough detuning mechanism in the uniform KV beam, the distorted KV beam passes through the up threshold 4th order stop band and gets into the 3rd order stop band, Fig.~\ref{fig23:subfig:b}.  The emittance growth after period 100 lies in the 3rd order structure resonance. The beam profile in phase space turns into a triangular shape and can be  clearly identified in Fig.~\ref{fig25:subfig:a}. In this particular simulation, until period 200 the beam is still trapped in the 3rd order structure resonance stop band; it will be self-adapted and de-tuned out of the stop band to a safe region ultimately with significant emittance growth and beam halo generation if extended to 600 or  more FD periods. The emittance splitting in different degrees of freedom is a characteristic of unstable time varying eigenmode $I_{j;k,l}(s)$ which grows exponentially.

For the case of non-uniform WB beam, the 4th order inner density perturbation already exists in the initial rms matched beam profile, thus the emittance growth takes place immediately,  Fig.~\ref{fig25:subfig:b}. However, the non-uniform WB beam can also de-tune the structure resonance more or less. Under these effects of these two mutual restraint mechanisms, the beam does not pass into the 3rd order stop band but stays in the region at the upper threshold of the 4th order structure resonance stop band in the following periods (Fig.~\ref{fig23:subfig:d}), where the beam reaches an ``quasi-equilibrium" state and evolves with slight emittance exchange between different degrees of freedom (Fig.~\ref{fig23:subfig:c}). Again, beam halo is generated in the stop band and evolves along the following periods. Compared to the KV initial case, the WB initial beam suffers less rms  emittance growth and beam halo.

Higher orders of structure resonances can be studied in a similar way \cite{4}. One example is the  fifth order structure resonance  $5\sigma\approx \Phi_{5;5,0}\approx 180^\circ$ with a condition  ($\sigma_0=80^{\circ}$, $\sigma=25^{\circ}$). Generally, the higher order structure resonances effects are weak, and  may be easily washed out by the ``density nonlinearity".  However, it is pointed out that higher order resonances exist for all rational n. When beams tunes are located in areas  free of the major resonances, beam will suffer from a slightly linear emittance growth, reflecting the the scattering from  higher orders resonances \cite{22}. Similar discussion will be shown in Sec. IV. A. 

Clearly, the principles (1, 3-7) discussed at the beginning of this section are interlaced and applied to give a basic understanding of the results in PIC simulations. As to principles 2 and 8, they are discussed in recent research Ref.~\cite{16} in detail. The principle 9 is studied separately on halo mechanism.

\section{Beam revolution in the space charge limit $\sigma\approx 0^{\circ}$}
\begin{figure*}
    \centering
    \subfloat[\label{fig28:subfig:b}]{
        \includegraphics[width=.4\linewidth]{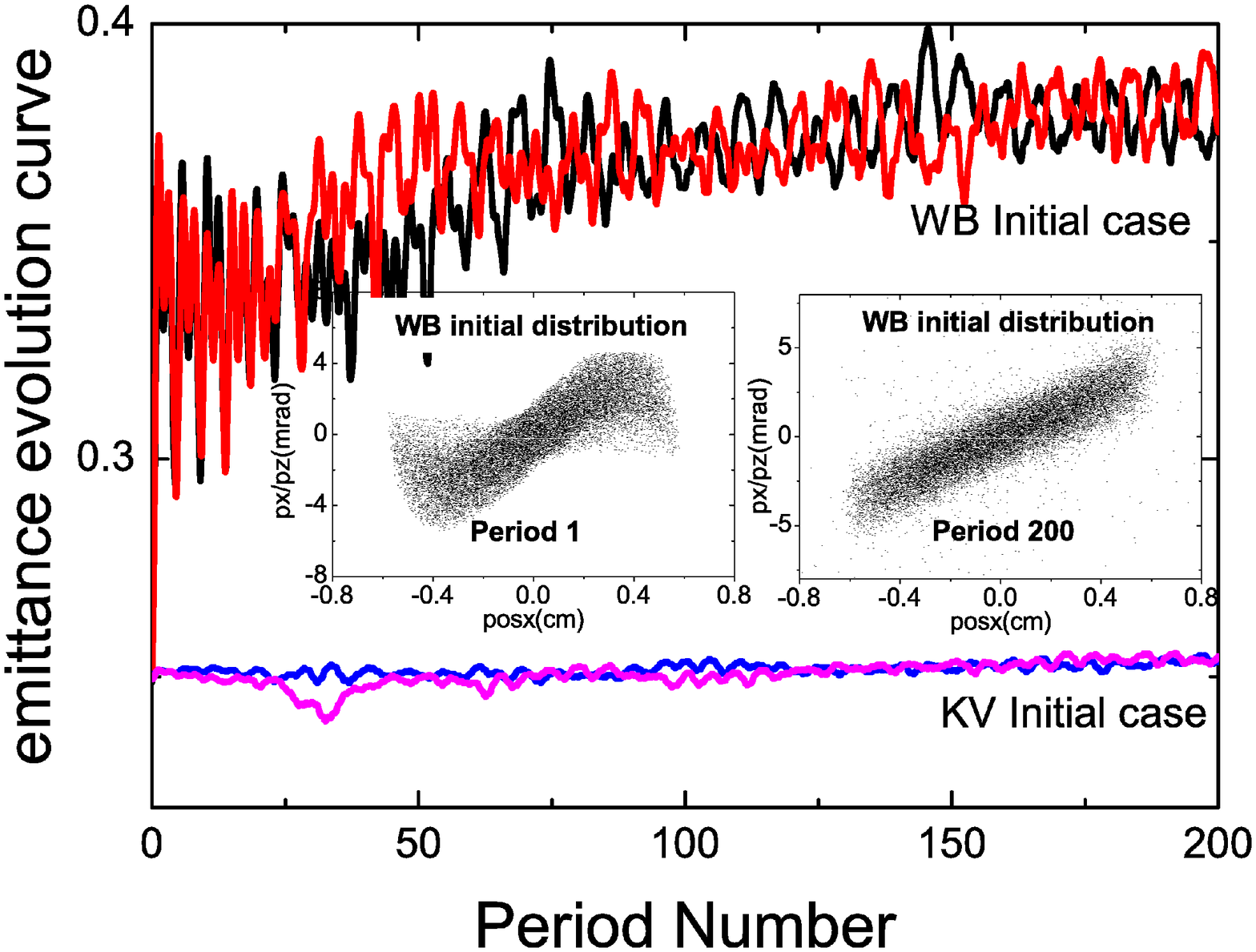}
    }
    \subfloat[\label{fig28:subfig:c}]{
        \includegraphics[width=.42\linewidth]{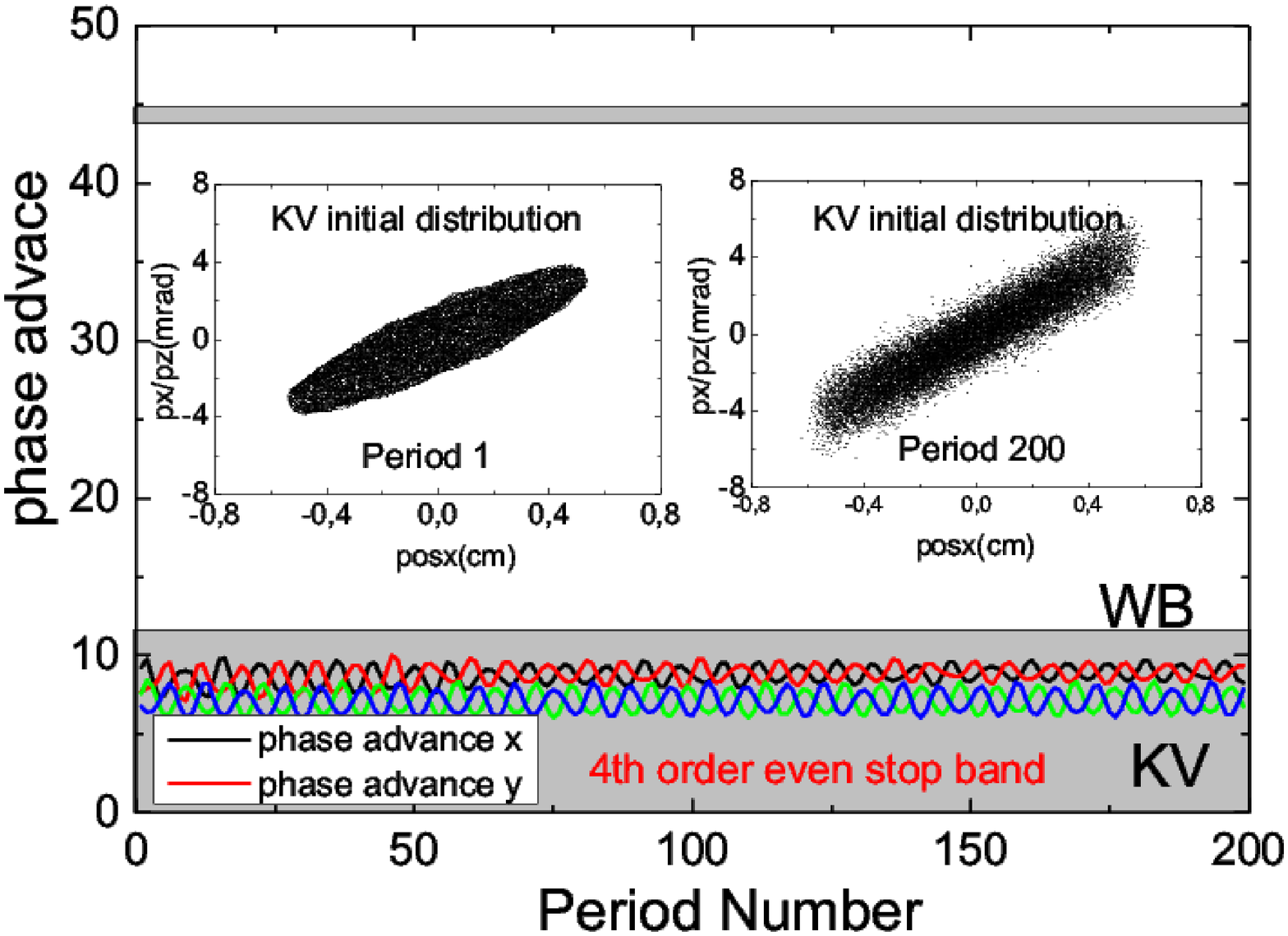}
    }
    \caption{\label{fig:fig28} The beam evolution near the space charge limit ($\sigma\approx 7^{\circ}$, $\sigma_0=50^{\circ}$) when the beam   initial locates in the mixed structure resonance stop band $n\sigma\approx\Phi_{j;k,l}=0^{\circ}$. }
\end{figure*}

\begin{figure*}
    \centering
    \subfloat[\label{fig33:subfig:a}]{
        \includegraphics[width=.17\linewidth]{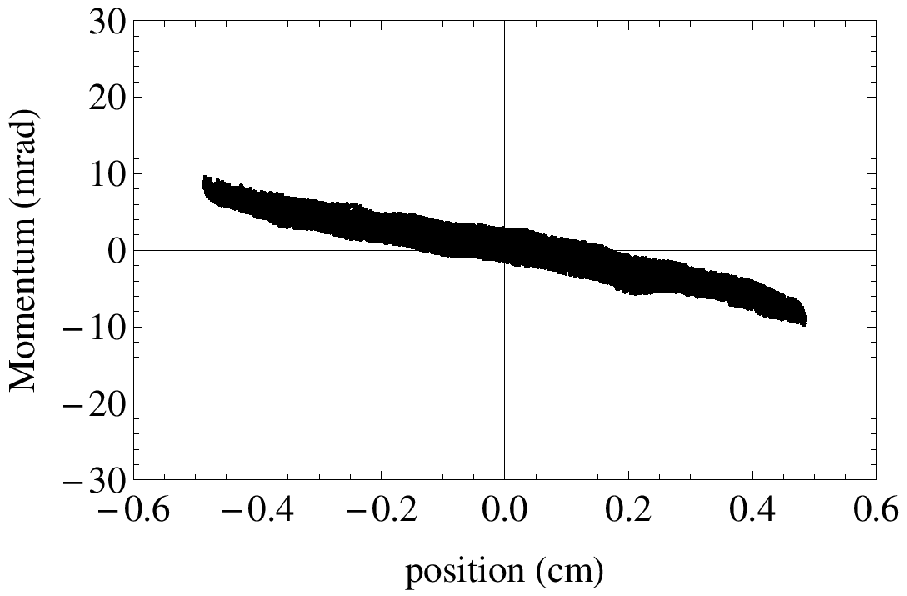}
    }
    \subfloat[\label{fig33:subfig:b}]{
        \includegraphics[width=.17\linewidth]{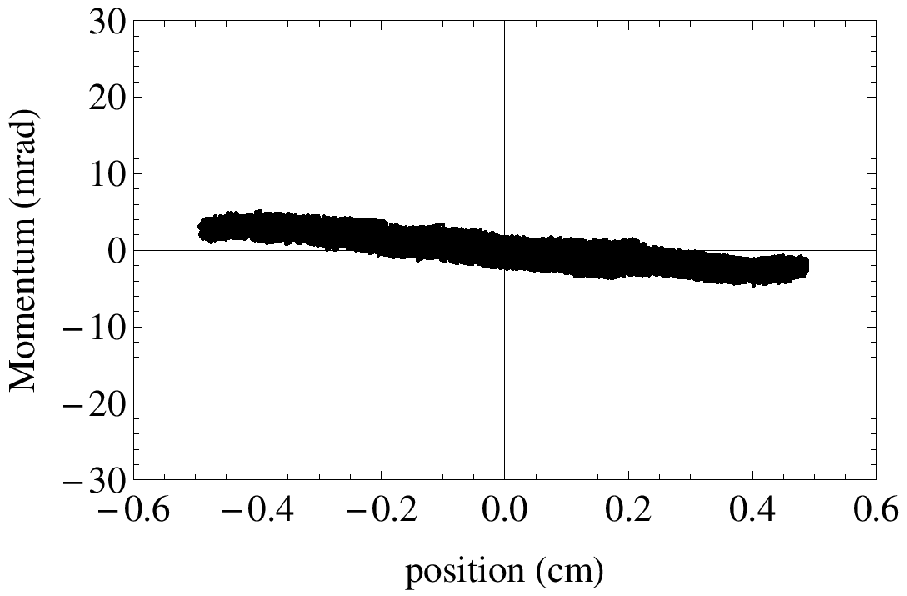}
    }
   \subfloat[\label{fig33:subfig:c}]{
        \includegraphics[width=.17\linewidth]{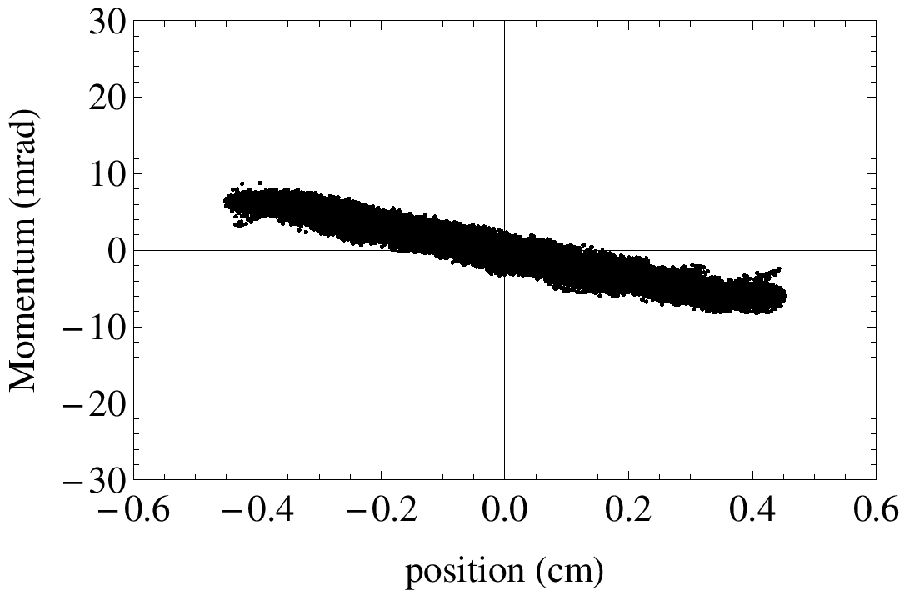}
    }
    \subfloat[\label{fig33:subfig:d}]{
        \includegraphics[width=.17\linewidth]{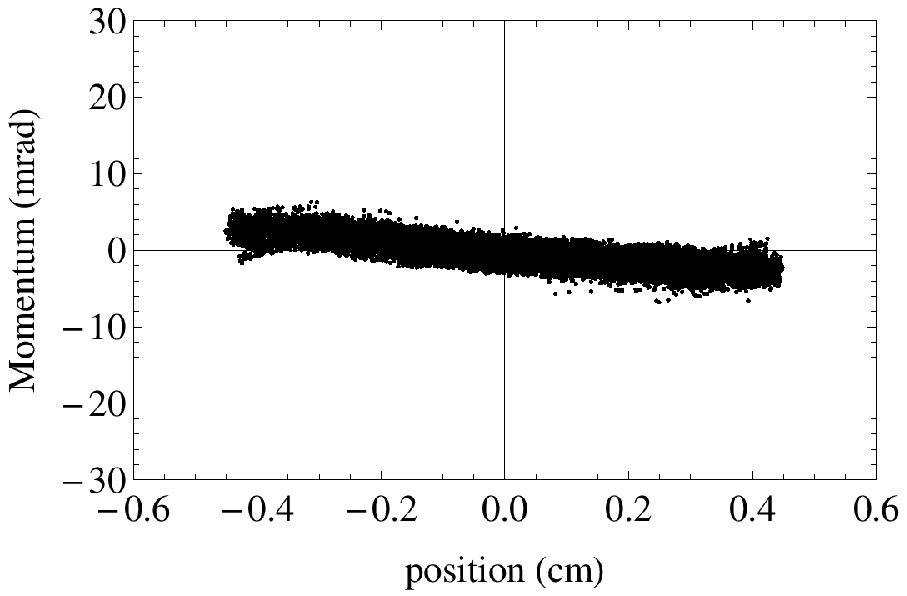}
    }
    \subfloat[\label{fig33:subfig:e}]{
        \includegraphics[width=.17\linewidth]{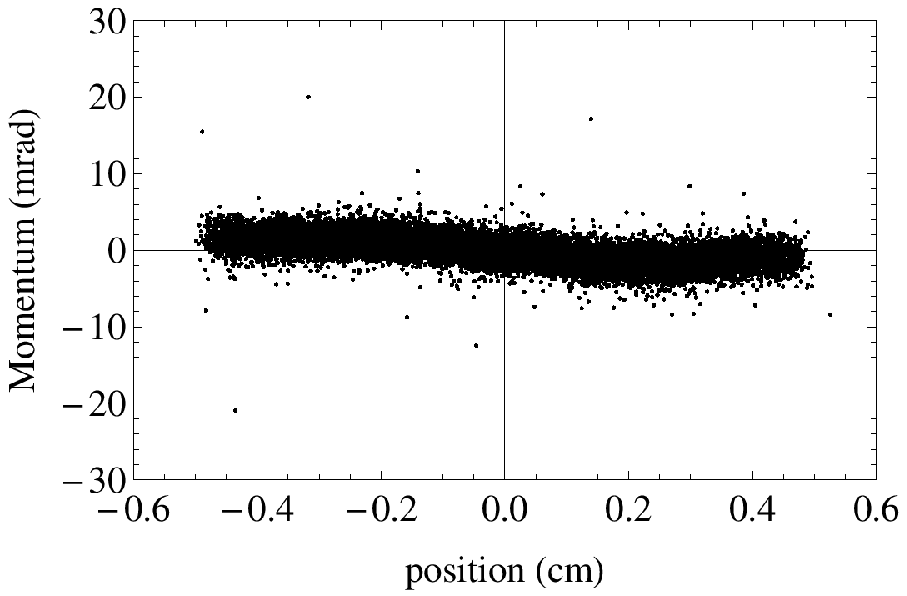}
    }
    
    \subfloat[\label{fig33:subfig:f}]{
        \includegraphics[width=.17\linewidth]{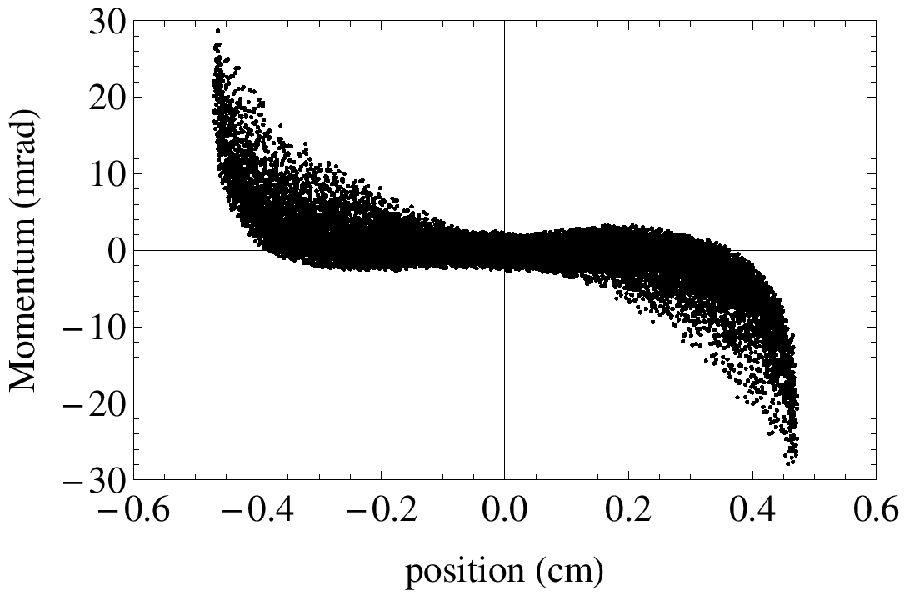}
    }
   \subfloat[\label{fig33:subfig:g}]{
        \includegraphics[width=.17\linewidth]{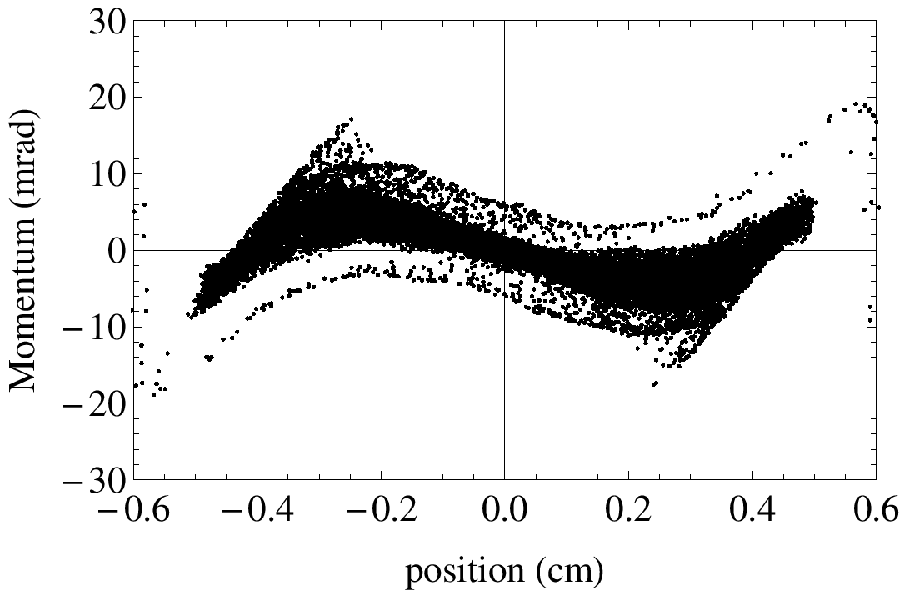}
    }
    \subfloat[\label{fig33:subfig:h}]{
        \includegraphics[width=.17\linewidth]{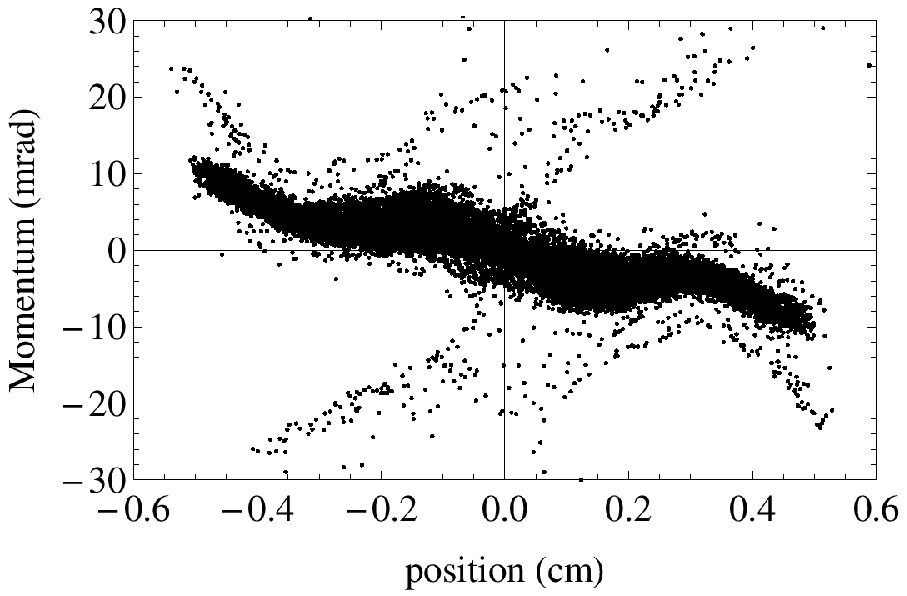}
    }
    \subfloat[\label{fig33:subfig:i}]{
        \includegraphics[width=.17\linewidth]{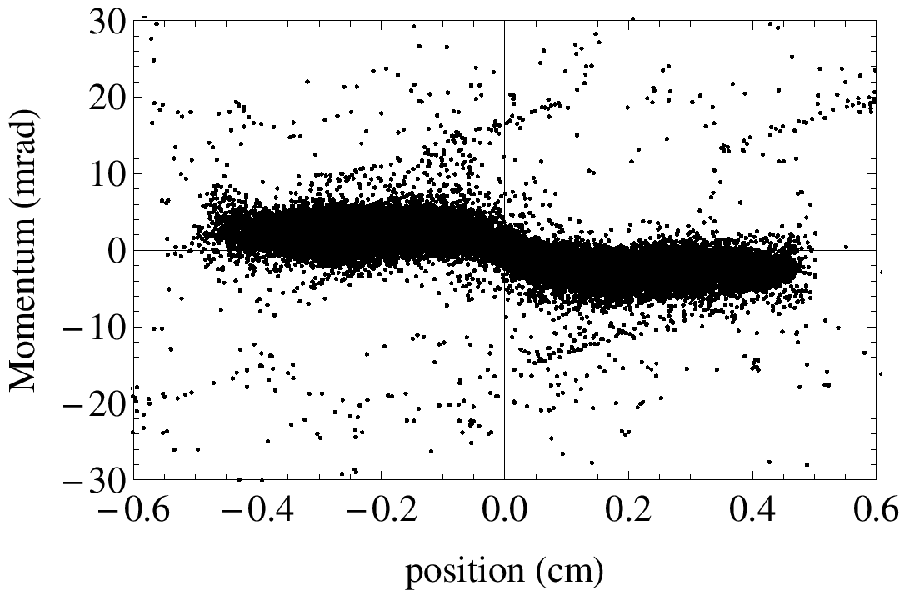}
    }
    \subfloat[\label{fig33:subfig:j}]{
        \includegraphics[width=.17\linewidth]{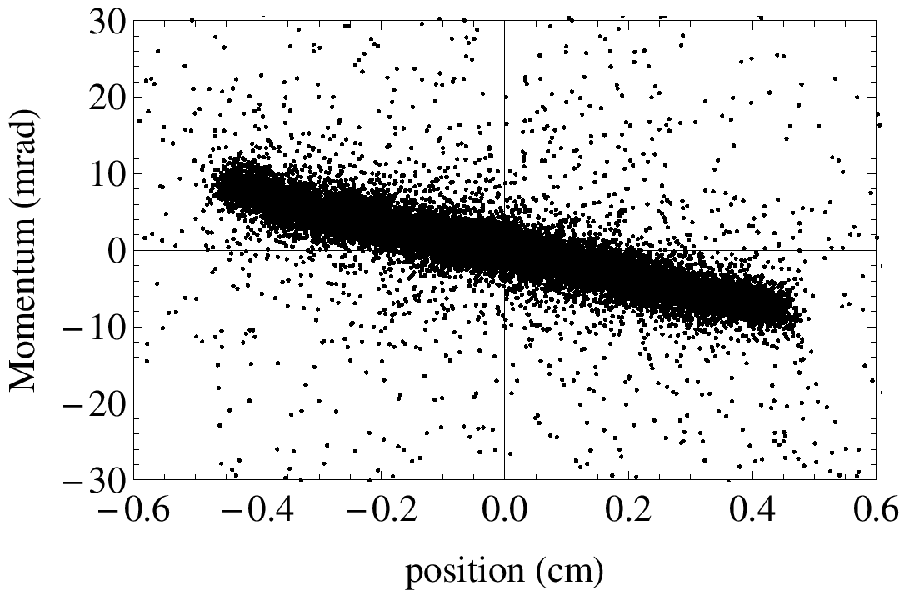}
    }
    \caption{\label{fig:fig33}Phase space profile evolution with KV initial beam (above) and  WB initial beam (below)  at period 1, 5, 12, 21, 191, left to right. The beam initial condition is $\sigma_0=80^{\circ}, \sigma=3^{\circ}$. }
\end{figure*}

In this section, the beam evolution in the space charge limit region $\sigma \approx 0^{\circ}$ will be discussed, where high orders structure resonances $n\sigma\approx\Phi_{j;k,l}=0^{\circ}$ can be excited and actually mixed together. In the following study, the stop band obtained from the 4th order structure resonance $\Phi_{4;0,4}=0^{\circ}$ is used to  represent the mixed structure resonance stop band roughly. There are two mechanisms contributing to the transient beam state: one is the resonant effect and the other is non-resonant effect. The resonant mechanism is due to the scattering from different orders of structure resonances, and the non-resonant effect is due to the non-neutral plasma characteristic of beam.

\subsection{Beam interaction with the structure resonance $n\sigma\approx\Phi_{j;k,l}= 0^{\circ}$ -- resonant effect}
In the former study \cite{1}, it is proved that the structure resonance will be degenerated to the resonance studied by Hofmann if smoothing approximation is applied. It noteworthy that these structure resonances near the space charge limit  when $n\sigma\approx\Phi_{j;k,l}= 0^{\circ}$ are exactly the same as those effective unstable ``non-oscillatory modes"  in the smoothed channel \cite{2}. In the AG channel discussed here,  numerical simulations are  launched with the condition as ($\sigma_0=50^{\circ}$, $\sigma=5^{\circ}$), ($\sigma_0=50^{\circ}$, $\sigma=7^{\circ}$) and ($\sigma_0=80^{\circ}$, $\sigma=3^{\circ}$).  There are no qualitative discrepancies among them. Here, the initial condition as  ($\sigma_0=50^{\circ}$, $\sigma=7^{\circ}$) is chosen to show the interaction between  beam and the structure resonance  $\Phi_{j;k,l}=0^{\circ}$. 

Figure~\ref{fig:fig28} shows  how the phase advance and rms emittance evolve with ($\sigma_0= 50^{\circ}$, $\sigma= 7^{\circ}$). The grey region in Fig.~\ref{fig28:subfig:c} near $\sigma= 0^{\circ}$ (decided by the condition $\Phi_{4;0,4}= 0^{\circ}$) is used to represent the mixed structure resonance stop band $n\sigma\approx\Phi_{j;k,l}= 0^{\circ}$.  Firstly, the rms emittance for the WB initial beam case has a quick growth at the very first several FD periods, leading to a separation of the starting points in the curves of phase advance. This initial emittance growth accompanied with a ``S-shaping" in phase space (also termed as ``particle redistribution" \cite{7}) is explained in the following sections). In addition, there exists a  linear emittance growth tendency both in  KV and WB beams which is  understood as a scattering effect between beam and overlapped   structure resonances. The settling time of this scattering between structure resonance and beam can be infinite long to make beam reaching the final equilibrium state. In the simulation within 200 periods, neither the KV nor the WB beam can get out of the stop band spontaneously. However, depending on the transient  parameters those the  beam undergoes, the beam  would show specific $n-fold$ phase space structure during the evolution. Fig.~(\ref{fig:fig33})  shows the beam phase space profile evolution of a KV and WB initial beam distribution when $\sigma=3^{\circ}, \sigma_0=80^{\circ}$. At certain period, beam phase space shows the influence of the higher order modes. 

\subsection{Plasma period, beam redistribution, free field energy limit and s-shaping -- non-resonant effect}

Considering the high intensity beam as a non-neutral plasma, the plasma period $\omega_p$  depicts one oscillation of the interdependence between beam density and energy modulation. During the time scale of $1/4$ plasma period, the density modulation reaches a maximum after a propagation distance corresponding to a $\pi/2$ plasma phase advance, concurrently with a minimum in energy modulation \cite{42}. Clearly, the plasma oscillation with the frequency $\omega_p$ is an  intrinsic characteristic in beam evolution, no matter if the beam is in the structure resonance stop band or not. It is reasonable to state that the ``phase phase mixing",  lying in this transient plasma oscillation when space charge is taken into account, is the mechanism that  ties the resonant effects, non-resonant effects and macroscopic beam phenomena together.

The phenomena of beam redistribution, ``free field energy limit", and S-shaping have been discussed for years \cite{7} and here we attribute them as  non-resonant effects. With the rms matched beam condition in AG focusing channels, the space charge field energy discrepancies between the beam  used and the stationary equilibrium state beam (here KV beam as the known stationary distribution function is used for estimation mathematically) is termed as ``free field energy" \cite{19,20}. The free field energy can be partially released in the time scale of $1/4$ plasma period, accompanying with rms emittance growth, known as the process of particle redistribution. The free field energy limit is used to estimate the maximum emittance growth, when the free field energy that a beam could have  been totally released. The maximum rms emittance growth is estimated 
\begin{eqnarray}\label{eq6.1}
\frac{\epsilon_{max}}{\epsilon}=\sqrt{1+a((\frac{\sigma_0}{\sigma})^2-1)}
\end{eqnarray}
where the coefficient $a$ is different according to the  beam distribution function used. The number of focusing periods that the 1/4 plasma period lasts can be estimated
\begin{eqnarray}\label{eq6.2}
N_{\frac{1}{4}\omega_p}\approx 90^{\circ}/\sqrt{2(\sigma_0^2-\sigma^2)}
\end{eqnarray}

Compared to the influence from resonant effects, which always need a certain time to develop to show their effects, the process of these non-resonant effects take place in the time scale of $1/4$ plasma period -- a quick manner. The time scales of resonant and non-resonant phenomena are the keys if one tries to distinguish the reasons for rms emittance growth in numerical simulations. It is noteworthy that the reason for ``S-shaping" in phase space in the space charge limit region $\sigma\approx 0^{\circ}$ should not be confused with the 2nd order structure resonance. Here the 2-fold phase space structure appearing with WB initial beam condition at the very first period, Fig.~\ref{fig33:subfig:f}, is due to a strong aberration caused by the nonlinar space charge.

The tune depression $\eta=\sigma/\sigma_0\approx 0.4$ has been  considered as a sort of space charge limit, where the various structure resonances (mainly  $n=0$) can be excited in most of the cases \cite{17}. However, according to the study discussed here, besides the effects from non-resonant plasma oscillation, we  make conclusion that the mixture structure resonances  $\Phi_{j;k,l}=0^{\circ}$ near the space charge limit  $\eta=0$ or higher order resonances anywhere on the transient phase advance trajectory can affect the beam but only lead to a small emittance growth tendency due to structure resonance scattering. It is reasonable to believe that the more density nonlinarity the beam brings, the less resonant effects are foreseen.  For a general high intensity accelerator design, instead of this 0.4 criteria, we purpose that current limit in beam dynamics should be decided as a compromise within the comprehensive consideration,  as how smooth is the  focusing lattice, structure resonance conditions, loss toleration, etc..

\section{Summary and discussions}
In this paper, we summarized the basic principles of the  interaction between high intensity beam and structure resonances. The non-resonant effect of space charge caused phase space mixing on  particle redistribution and ``free field limit" are also revisited. The right understanding of these effects will lead to a clearer understanding of the space charge beam physics.      

Although the Vlasov-Poisson model predicts the phenomena found in simulation very well, still because of  the  assumptions and techniques used in the theoretical part, it can not answer  all of the questions about beam collective motions due to space charge, even when the beam is limited to 2D degrees of freedom.  The fundamental question comes from the non-physical KV initial beam assumption and the linearized perturbation techniques. The beam behaviour has to be studied in a transient sense with  time dependent varying parameters, which actually violent the basic assumption immediately. Thus the effects from the nonlinear-detuning and Landau damping are actually not included in theory. Still, the resonance studies obtained from linearized perturbation  theory are still very useful when they are used to investigate the  transient state as  a guidance to get a better beam dynamic behavior in accelerators.   

\section{Acknowledgements }
Great thanks for helpful discussion with H. Okamoto in Hiroshima University. This work is supported by the Key Research Program of Frontier Sciences CAS (QYZDJ-SSW-SLH001) and NSFC (11775239).

\bibliography{reference}


\end{document}